\documentclass{ws-p8-50x6-00}
\usepackage{wrapfig}
\usepackage{rotating}

\begin{document}

\title {Overview of Relativistic Heavy-Ion Physics}
\author{Itzhak Tserruya}
\address{Dept. of Particle Physics, Weizmann Institute, Rehovot 76100, Israel \\ 
        E-mail: Itzhak.Tserruya@weizmann.ac.il}
\maketitle

\abstracts{
Selected topics in the field of relativistic heavy-ion collisions are reviewed from
the 15 year research programme at the SPS at CERN  and the AGS at BNL,  and from the first 
run of the Relativistic Heavy-Ion Collider at BNL.}

\section{Introduction}
 The Relativistic Heavy-Ion Collider (RHIC) at BNL started regular operation in the
summer of year 2000 opening new horizons in the study of nuclear matter under 
extreme conditions of pressure, density and temperature. The primary scientific 
objective is the search for the phase transition associated with quark-gluon plasma 
formation and chiral symmetry  restoration, predicted to occur under these conditions. 
According to QCD calculations on the lattice, with three light quark flavors  
(two degenerate u,d-quark masses and a strange quark), the phase transition 
should be of first order \cite{karsch-qm01}.
  
   In a very short, but extremely successful run with an integrated luminosity of 
only a few $\mu b^{-1}$, the four RHIC experiments have delivered an impressive amount 
of results offering a first glimpse at the physics of relativistic heavy-ion 
collisions at the highest energies achieved to date, $\sqrt{s_{NN}}$ = 130 GeV, 
almost one order of magnitude larger than the highest energy of $\sqrt{s_{NN}}$ = 17.2 GeV 
available at CERN. Together with that, the 15 years of relativistic heavy-ion physics at 
the CERN SPS and the BNL AGS have produced a wealth of very interesting and intriguing 
results in the quest for the quark-gluon plasma. 

 About a dozen of different experiments at the AGS and SPS and four experiments at RHIC are or 
have been involved in this endeavour covering a very broad range of observables which can be 
divided into two broad categories:
(i) global  and  hadronic observables (transverse energy and charged particle distributions, 
$p_t$ spectra, particle production rates, two-particle correlations, flow, etc.); they provide crucial 
information about the reaction dynamics and in particular about the the size and properties of 
the final hadronic system at freeze-out i.e. when hadrons cease to interact. From their 
systematic studies, the picture of a chemically and thermally equilibrated hadronic system undergoing  
collective expansion has emerged, with freeze-out parameters (in the temperature vs. baryo-chemical 
potential plane) close to the expected phase transition boudaries. 
(ii) observables which have been proposed as signatures for the phase transition:
the most prominent ones for deconfinement are  $J/\psi$ suppression, real and virtual thermal photons
and jet quenching. Low-mass dileptons and the $\rho$ meson  with its very short lifetime, 
are considered  the best probes for the chiral symmetry restoration phase transition. 	

   In the limited space  of this paper it is not possible to do justice to the vast amount of
available results \cite{qm01}. A selection is therefore unavoidable and this paper is restricted to 
selected topics on (i) global observables where systematic comparisons are made from AGS up to RHIC 
energies, (ii) the observation of $J/\psi$ suppression and excess emission of low-mass lepton 
pairs which are among the most notable results hinting at new physics from the 
SPS programme (results on these two topics are not yet available at RHIC as they require 
a much higher luminosity than achieved in the first year), and (iii) the suppression of high 
p$_T$ hadrons which is one of the highlights of the first RHIC run pointing to new phenomena
opening up at the high energies of RHIC.

\vspace{-0.2cm}
\section{Global Observables}
\vspace{-0.2cm}
   Global observables, like multiplicity and transverse energy, 
provide very valuable information. Besides defining the collision geometry, they 
can be related to the initial energy density, e.g. using the well-known Bjorken 
relation \cite{bjorken}, shed light into the mechanisms of particle production\cite{wang-miklos}
and provide constraints to the many models aiming at describing these collisions.  

   The charged particle rapidity density at mid-rapidity has been measured by the four 
RHIC experiments -BRAHMS, PHENIX, PHOBOS and STAR- with very good agreement among 
their results \cite{phobos-200}. For central Au-Au collisions at $\sqrt{s_{NN}}$ = 130 GeV, the global 
average is $dN_{ch}/d\eta = 580 \pm 18$. For the transverse energy rapidity density there is
only one measurement by PHENIX, with a value $dE_T/d\eta = 578 ^{+26}_{-39}$ GeV  for the most central
2\% of the inelastic cross section \cite{phenix-ppg002}. Using the Bjorken formula \cite{bjorken} this 
translates into an initial energy density of $\epsilon$ = 5.0 GeV/fm$^3$ \footnote{if one uses the 
canonical formation time $\tau$ = 1fm/c. Much shorter formation times as advocated in saturation models
\cite{kharzeev-nardi-0012025} would result in energy densities as high as  
$\epsilon \simeq 20$ GeV/fm$^3$.}. These values are 60-70\% larger than the corresponding ones at the 
full SPS energy.

The energy dependence of the charged particle density $dN_{ch}/d\eta$, normalized per pair of participating 
nucleons (N$_{part}/2$), exhibits a boring logarithmic rise with  $\sqrt{s_{NN}}$  from AGS up to RHIC energies, 
as shown in Fig.~1. However, the dependence is very different from the $p{\overline p}$ systematics also shown 
in Fig.~1. At $\sqrt{s_{NN}}$ = 200 GeV, $\sim$65\%  more particles per pair of participants are produced 
in Au-Au collisions than in $p{\overline p}$. I shall return to this point below.
Note also that the large increase predicted by the HIJING model with jet quenching \cite{wang-miklos}
(upper curve in Fig. 1) is not observed, and from $\sqrt{s_{NN}}$ = 130 to 200 GeV the particle density 
increases only by 15\%  
\cite{phobos-200}.   
%%%%% FIGURE 1 SQRT(S) DEPENDENCE OF PARTICLE DENSITY
\begin{wrapfigure}[14]{r}{5.0cm}
\epsfxsize=11pc                   % will enlarge or reduce the postscript figures based on the xsize
\vspace*{-0.3cm}
\epsfbox{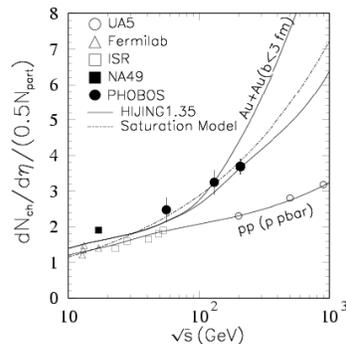}   
\vspace*{-0.5cm}
\caption {Energy dependence of charged particle density in central A-A and pp collisions
~\protect\cite{wang-miklos}.}  
\end{wrapfigure} 
   
   The centrality dependence of the charged particle density has been proposed as a sensitive
tool to shed light on the particle production mechanism: whereas soft processes  are believed 
to scale with the number of participants N$_{part}$, hard processes, expected to play a significant 
role as the energy increases, lead to a scaling with the number of binary collisions N$_{bin}$. 
Indeed, PHENIX and PHOBOS have reported an increase of  $dN_{ch}/d\eta$ and $dE_T/d\eta$
strongly than linear with  N$_{part}$ (the PHENIX results are shown in
Fig. 2 \cite{phenix-ppg001}) and in the framework of models with these two components, like 
HIJING \cite{kharzeev-nardi-0012025,hijing}, such an increase is interpreted as evidence of 
the contribution of hard processes to particle production \footnote {This is, however, not a 
unique interpretation. The centrality dependence results of PHENIX shown in Fig. 2 have also 
been explained with a model based on purely soft processes \cite{capella-sousa-0101023}.}. This 
is to be contrasted to the results obtained at the SPS where WA98 reports a much weaker increase 
(also shown in Fig. 2) \cite{wa98-cent} and WA97 an even weaker increase consistent within errors 
with proportionality between multiplicity and participants \cite{wa97-cent}. 
Both the PHENIX and the WA98 data, extrapolated to peripheral collisions, are in very 
good agreement with the $p{\overline p}$ result at the same $\sqrt{s}$ derived from the UA5
data \cite{ua5}. The increased role of hard processes at higher energies
and more central collisions would then be the explanation for the faster increase of particle 
production in A-A collisions compared to  $p{\overline p}$ previously discussed in the context 
of Fig. 1. The importance of hard processes at RHIC energies is a very interesting issue which 
will be further discussed in this review.

   More surprising is the behavior of the ratio  $dE_T/d\eta / dN_{ch}/d\eta$, the average transverse 
energy per charged particle, shown in Fig. 3. 
This ratio is found to be independent of centrality  and approximately equal to $\sim$0.8 GeV.
Within errors of the order of 10-20\%  this ratio is also independent 
of $\sqrt{s_{NN}}$ from AGS up to RHIC, implying a constant or a very moderate increase of the 
average p$_T$ per particle. It is also interesting to note that UA1 quotes a very similar ratio 
at $\sqrt{s}$ = 200 GeV \cite{ua1}. This seemingly universal behaviour of constant energy/particle 
is one of the most puzzling results. The increase in $dE_T/d\eta$ with  $\sqrt{s_{NN}}$  translates 
into an increase in the number of produced particles rather than in the production of particles 
with higher $E_T$. 
%%%%%% FIGURE 2 AND 3 SIDE BY SIDE
%%%%%% FIGURE 2 CENTRALITY DEPENDENCE OF CHARGED PARTICLE PRODUCTION
%%% I may need to remove the vertical space of -1.0cm before the begining of the figure. It pushes
%%% the figure up in the page
\begin{figure}[t]
%\vspace{-0.2cm}
\begin{minipage}[t]{55mm}
\epsfig{file=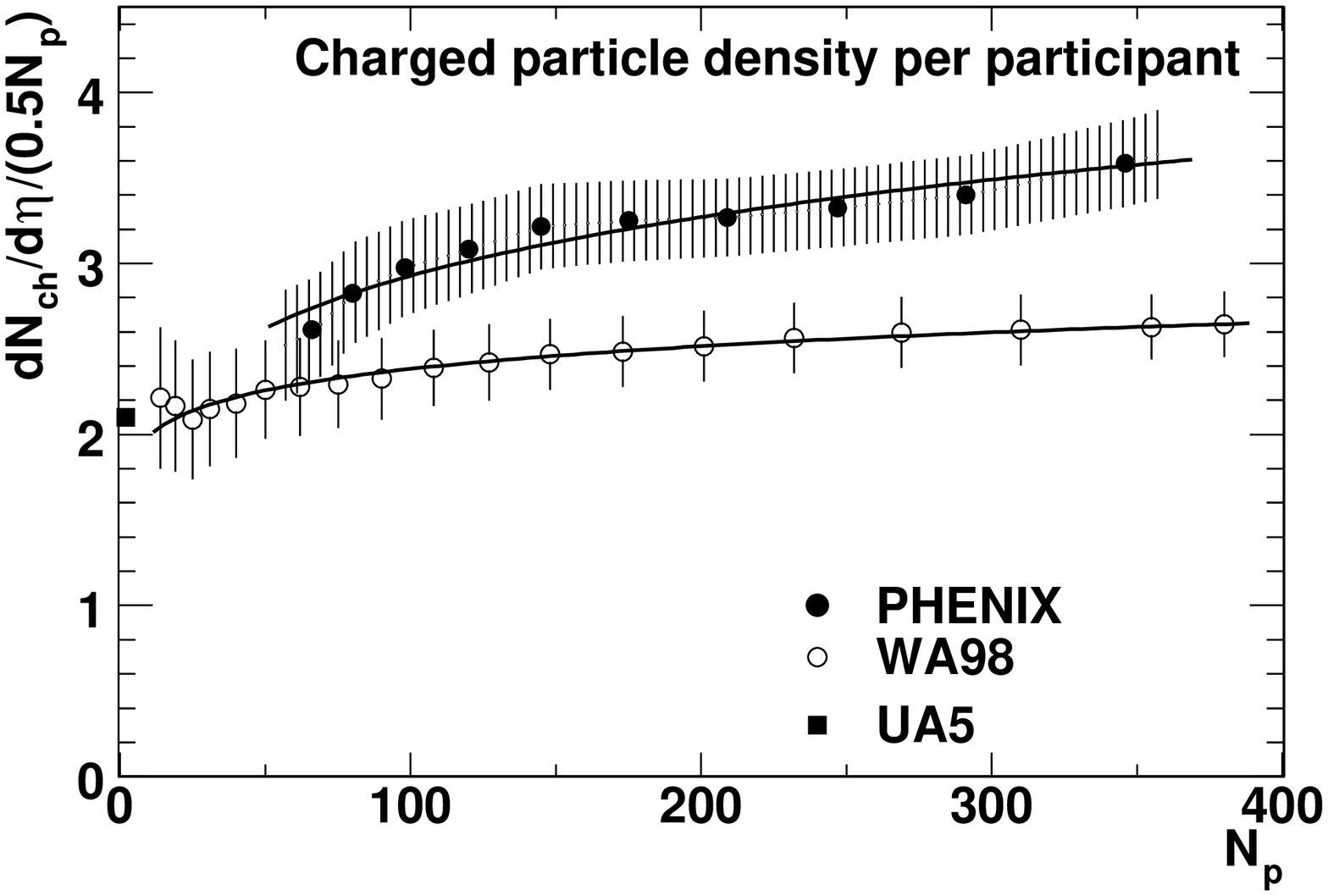,width=5.5cm,height=5.0cm}
\vspace{-1.0cm}
\caption{Centrality dependence of the charged particle density from
PHENIX and WA98 ~\protect\cite{phenix-qm01-milov}.}
\end{minipage}
\hspace{\fill}
%%%%%%% FIGURE 3.  ET PER CHARGED PARTICLE VS CENTRALITY 
\begin{minipage}[ht]{55mm}
\vspace{-4.3cm}
%\hspace{5.5cm}
%if I put a vspace here it produces two figures one below the other
\epsfig{file=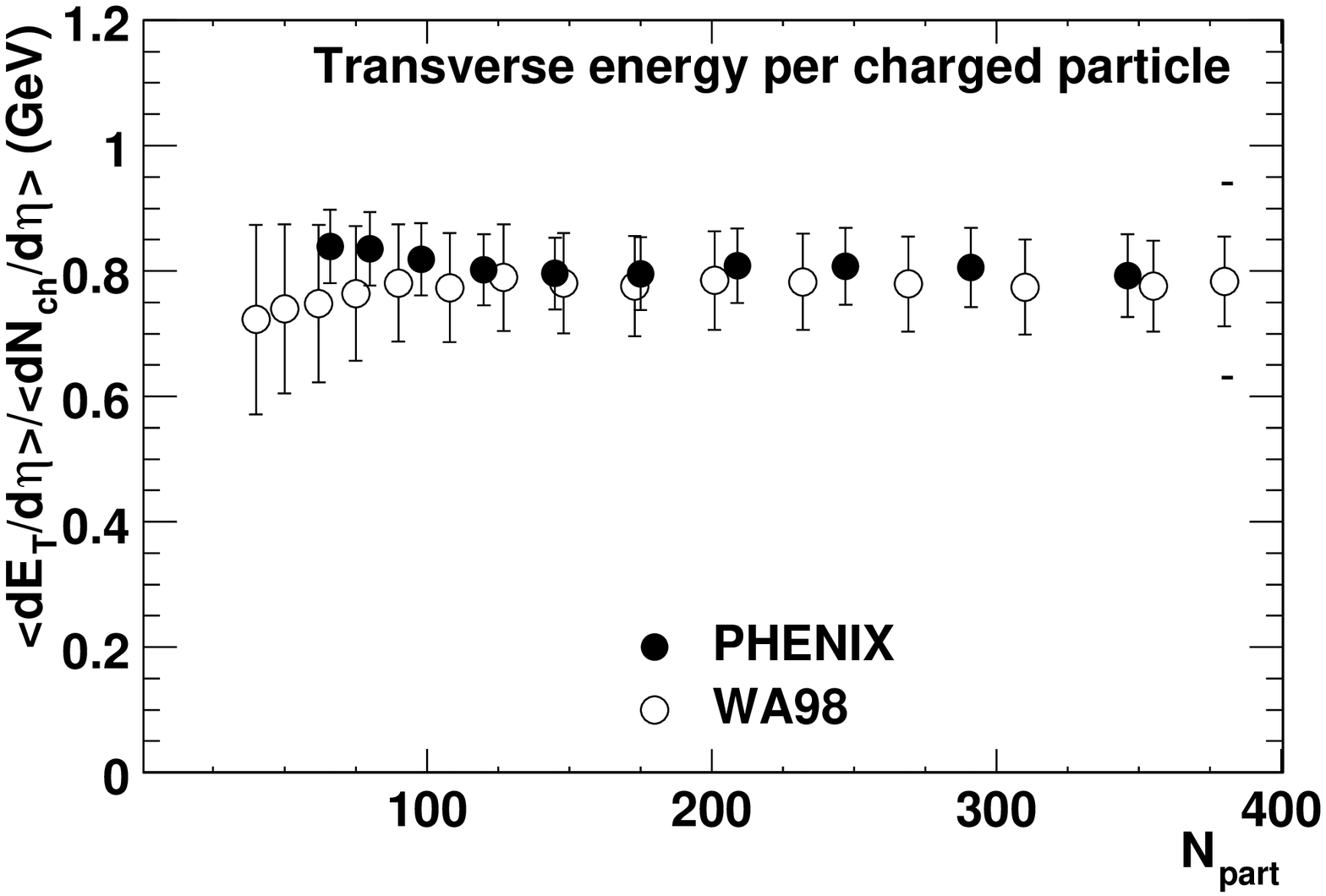,height=5.0cm,width=5.5cm}
\vspace{-1.0cm}
\caption{Average $E_T$ per charged particle from PHENIX and WA98 ~\protect\cite{phenix-ppg002}.}
\end{minipage}
\vspace{-0.5cm}
\end{figure}
 
\section{Low-Mass $e^+e^-$ Pairs}

   Dileptons and photons are since a long time emphasized as unique probes to study
the dynamics of of relativistic heavy-ion collisions \cite{shuryak}.
 The interest stems from their relatively large mean free path. As a consequence, they
can leave the interaction region without final state interaction, carrying information about 
the conditions and properties of the matter at the time of their production
and in particular of the early stages of the collision when temperature and energy density have 
their largest values, i.e. when the conjectured deconfinement and chiral symmetry restoration phase 
transition has the best chance to occur. 

 The prominent topic of interest, both in dileptons and photons, 
is the identification of thermal radiation emitted from 
the collision system. This radiation should tell us the nature of the matter formed, 
a quark-gluon plasma (QGP) or a high-density hadron gas (HG). 
The physics potential of low-mass dileptons is further emphasized by their
sensitivity to chiral symmetry restoration. The $\rho$-meson is the prime agent here.
Due to its very short lifetime  ($\tau$ = 1.3 fm/c) compared to the typical fireball 
lifetime of $\sim$10 fm/c at SPS energies, most of the $\rho$ mesons 
decay inside the interaction region providing a unique 
opportunity to observe in-medium modifications of particle properties (mass and/or width)
which might be linked to chiral symmetry restoration. 
The situation is  different for the $\omega$ and $\phi$ mesons. Because 
of their much longer lifetimes they predominantly decay outside the interaction region after 
having regained their vacuum properties.
The  $\omega$ and $\phi$ mesons remain nevertheless important messengers: the 
undisturbed   $\omega$ can serve as a reference and the  $\phi$ with its $s{\overline s}$
content is a probe of strangeness production.

  The CERES experiment has systematically studied the production of low-mass $e^+e^-$ pairs, 
$m = 200 - 600 MeV/c^2$, with measurements of p-Be (a very good approximation to pp collisions) and
p-Au at 450 GeV/c \cite{pbe-ee,pbe-eegamma}, S-Au at 200~GeV per nucleon \cite{prl95},
and Pb-Au at 158 GeV per nucleon \cite{pbau95-plb,bl-qm99} and 
40~ GeV per nucleon \cite{pbau-40}.
Whereas the p data are well reproduced by the known hadronic sources, a strong enhancement of low-mass
pairs is observed both in the S and Pb data with respect to those sources, scaled to the nuclear case with
the event multiplicity. Fig. 4 shows the preliminary mass spectrum from Pb-Au collisions at 40 A GeV  
with an enhancement factor of $5.0\pm1.5(stat.)$ relative to the hadronic cocktail for masses $m > 200 MeV/c^2$.
With the large statistical uncertainty, this is consistent with the enancement factor of
$2.9\pm0.3(stat.)\pm0.6(syst.)$ obtained from the combined 95-96 Pb-Au data at 158 A GeV.
Further studies on the latter demonstrate that the enhancement is more pronounced at low pair p$_T$
and increases faster than linearly with the event multiplicity. In all cases, with the S and Pb beam, 
the enhancement sets at $m \geq 2m_{\pi}$. 
%%%% FIGURES 4 AND 5 SIDE BY SIDE
%%%% FIGURE 4 RIGHT CERES 40 A GEV MASS SPECTRUM   
%%% I may need to remove the vertical space of -1.0cm before the begining of the figure. It pushes
%%% the figure up in the page
\begin{figure}[t]
%\vspace{-0.3cm}  
\begin{minipage}[t]{55mm}
\epsfig{file=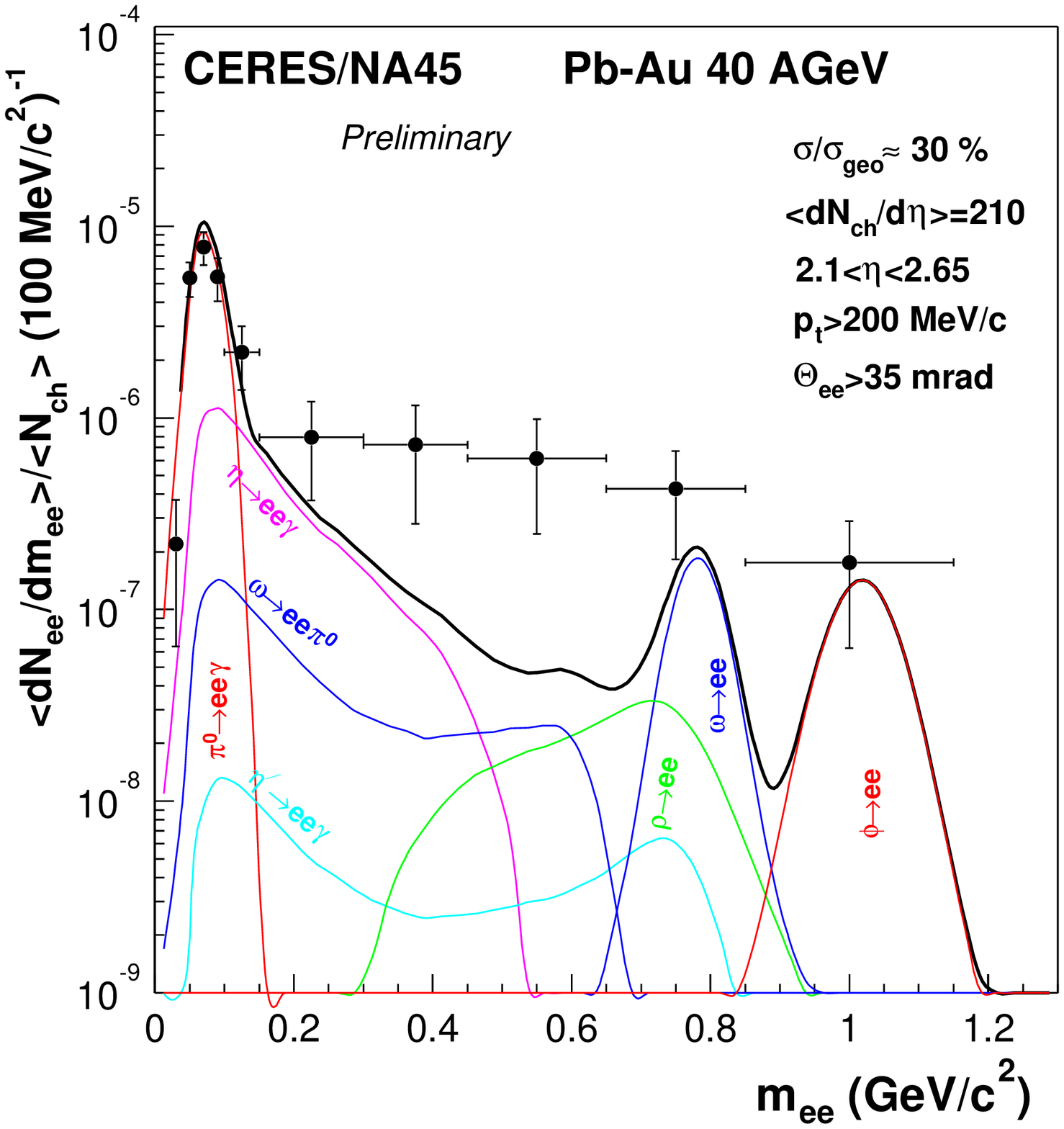,height=5.5cm}
\vspace{-1.0cm}
\caption{Invariant low-mass $e^+e^-$ spectrum measured by CERES in 40 A GeV~\protect\cite{pbau-40}
         Pb-Au collisions. The figure also shows the summed (thick solid line) and individual 
         contributions from hadronic sources.}
\end{minipage}
\hspace{\fill}
%%%%%%% FIGURE 5.   PB 96 WITH ALL MODELS 
\begin{minipage}[ht]{55mm}
\vspace{-4.3cm}
\begin{rotate}{-90}
\epsfig{file=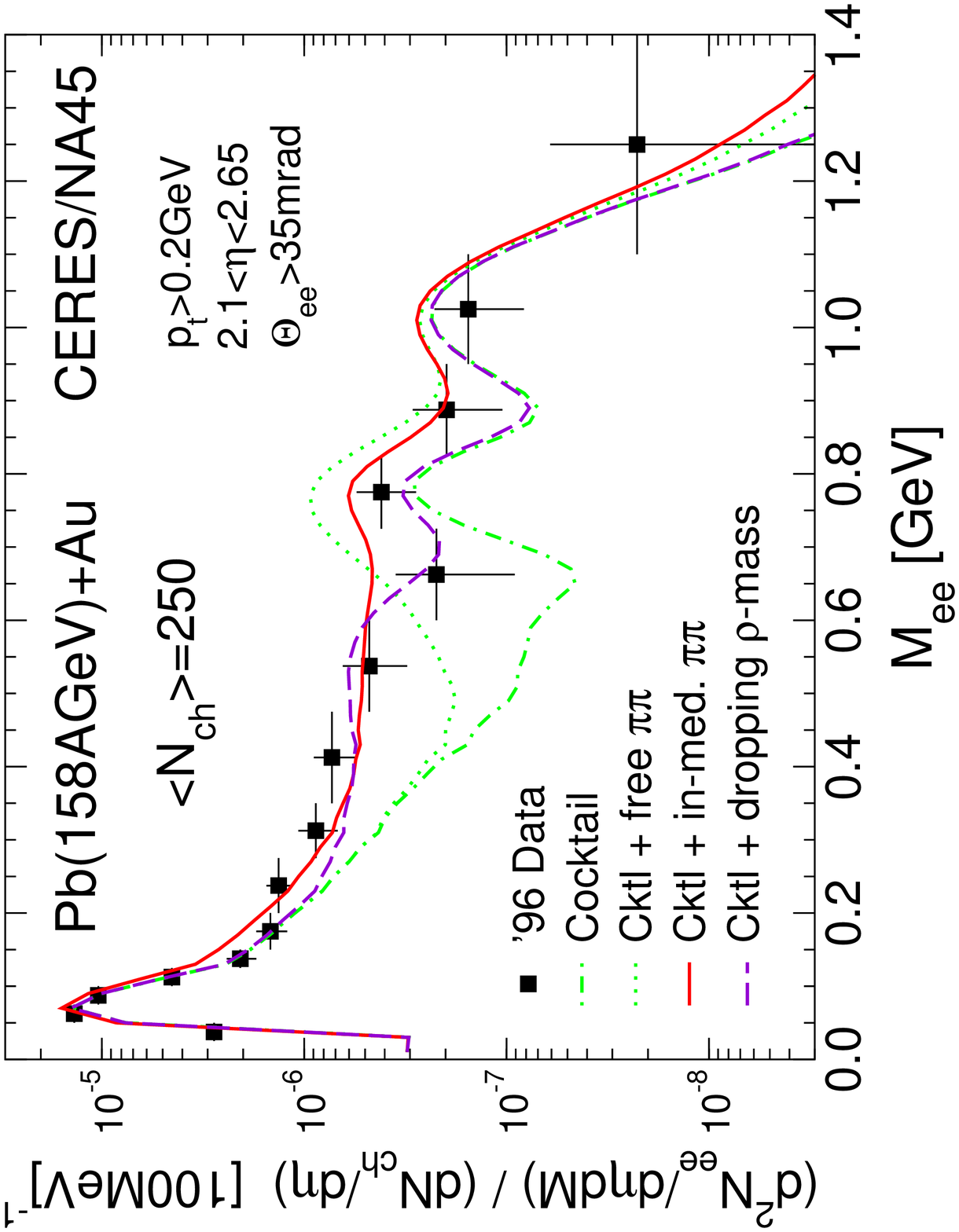,width=5.9cm,height=6.0cm}
\end{rotate}
\vspace{5.3cm}
\caption {Comparison of CERES 96 results from Pb-Au collisions at 158 A GeV with 
          various theoretical approaches (see text).}
\end{minipage}
\vspace{-0.5cm}
\end{figure}

  The enhancement of low-mass dileptons has triggered a wealth of theoretical activity. Dozens of 
articles have been published on the subject. (For a comprehensive review see \cite{rapp-wambach}).  
There is consensus that  a simple superposition of pp collisions cannot explain the data and that 
an additional source is needed. The pion annihilation channel ($\pi^+\pi^- \rightarrow \rho \rightarrow l^+l^-$), 
obviously irrelevant in pp collisions, has to be added in the nuclear case. This channel
accounts for a large fraction of the observed enhancement (see line cocktail + free $\pi\pi$ in Fig. 5)
and provides first evidence of thermal radiation from a dense hadron gas. However, in order to 
quantitatively reproduce the data in the mass region 0.2 $< m_{e^+e^-} <$ 0.6 GeV/c$^2$, it was 
found necessary to include in-medium modifications of the $\rho$ meson. Li, Ko and Brown \cite{li-ko-brown}, 
following the original Brown-Rho scaling \cite{brown-rho}, proposed a decrease of the $\rho$-mass in 
the hot and dense fireball, as a precursor of chiral symmetry restoration, and achieved excellent agreement 
with the CERES data (see the line cocktail + dropping mass in Fig. 5). 
 
 Another avenue, based on effective Lagrangians, uses the broadening of the $\rho$-meson spectral function 
resulting from its propagation in the medium, mainly from the scattering of $\rho$ mesons off baryons \cite{wambach}, 
and achieves also an excellent reproduction of the CERES data (see line cocktail + in-medium $\pi\pi$ in Fig. 5). 
The success of these two different approaches, one relying on quark degrees of freedom and the other on a 
pure hadronic model, has attracted considerable interest raising the hypothesis of quark-hadron duality. 
Rapp provided empirical support to this hypothesis by showing that in a high density hadron gas the dilepton 
production rates calculated with the in-medium  $\rho$-meson spectral function are very similar to the 
$q{\overline q}$ annihilation rates calculated in pQCD \cite{rapp-duality-qm99}.
 
Finally, it is interesting to note that the 40 A GeV data is also equally well reproduced by the 
dropping-mass and broadening scenarios \cite{pbau-40}. Much better data are needed, in order to 
discriminate among the two. It is hoped that the last CERES run of year 2000 taken with an improved 
mass resolution of $\sim$2\% at the $\omega$ mass and good statistics should allow that.

\section{$J/\psi$ Suppression}

  The melting of charmonium states in a deconfined state of matter is one of the
oldest signatures of QGP formation \cite{satz-matsui} and has provided one of the most
exciting sagas of the SPS programme. Already in the first runs with O and S beams, the
NA38/NA50 experiment observed a suppression of $J/\psi$ which immediately captured the interest 
\cite{na38-jpsi}. Intensive theoretical efforts were devoted to explain the suppression within 
conventional physics and it soon became clear that all experimental data, including systematic 
studies of pp, pA and S-U collisions, can be reproduced by invoking the absorption  in the nuclear
medium of the $c\overline{c}$ pair before it forms a $J/\psi$,  with an absorption cross 
section $\sigma_{abs} = 6.4 \pm 0.8$mb \cite{na38-jpsi}. 
%%%%%% FIGURE 6. Jpsi SUPPRESSION VS EPSILON
\begin{wrapfigure}[15]{r}{5cm}
\epsfxsize=12pc
%\vspace{-0.5cm}
\epsfbox{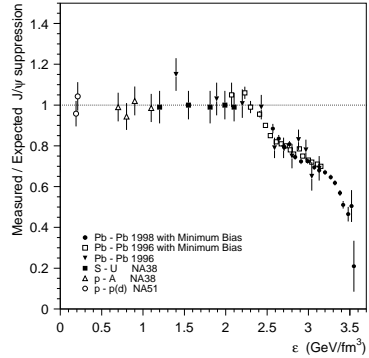}
\vspace{-0.9cm}           
 \caption{Measured over expected (assuming normal absorption cross section) J/$\psi$ yield vs. energy 
         density ~\protect\cite{na50-jpsi}.}
%\vspace{-0.8cm}
\end{wrapfigure}

   However, a different behavior is observed with the 158 A GeV Pb beam. Whereas peripheral 
collisions seem to  follow the regular absorption pattern, an anomalously larger suppression occurs 
at more central collisions characterized by impact parameters $b < 8 $fm or E$_T > 40 $GeV which 
can be translated into a Bjorken energy density $\epsilon > 2.3 $GeV/fm$^3$ (see Fig. 6 \cite{na50-jpsi}).
The data in this figure exhibit a two-step suppression pattern which NA50 attributes to the successive 
melting of the $\chi_c$ and directly produced J/$\psi$ mesons in a quark-gluon plasma scenario. 
  
The same data normalized to the Drell-Yan cross section is shown in Fig.~7 
as function of E$_T$. The two-step pattern can be discerned here at E$_T$ values of
$\sim 30$ and $\sim$ 100 GeV. Most published calculations, based on conventional hadronic models 
including the effect of absorption by comovers, fail to reproduce the results as illustrated in 
the left panel of Fig.~7. The recent calculations of Capella et al. \cite{capella} are in much better 
agreement with the data and fail only at the most central collisions (middle panel).
On the other hand, quite good agreement over the whole E$_T$ range is achieved by models assuming QGP formation
(right panel) \cite{blaizot2000,nardi-satz}. The model of Blaizot et al. reproduce the data remarkably well 
by invoking  $J/\psi$ suppression whenever the energy density exceeds the critical value for deconfinement 
(first step)  together with fluctuations of the transverse energy for the most central collisions (second step) 
\cite{blaizot2000}. 
%%%% FIGURES 7--  J/psi SUPPRESSION VS MODELS -- THREE PANELS SIDE BY SIDE ONE CAPTION
\begin{figure}[h]
\vspace{-1.0cm}
%%%% FIGURE  7a CONVENTIONAL HADRONIC MODELS
\includegraphics*[width=3.8cm]{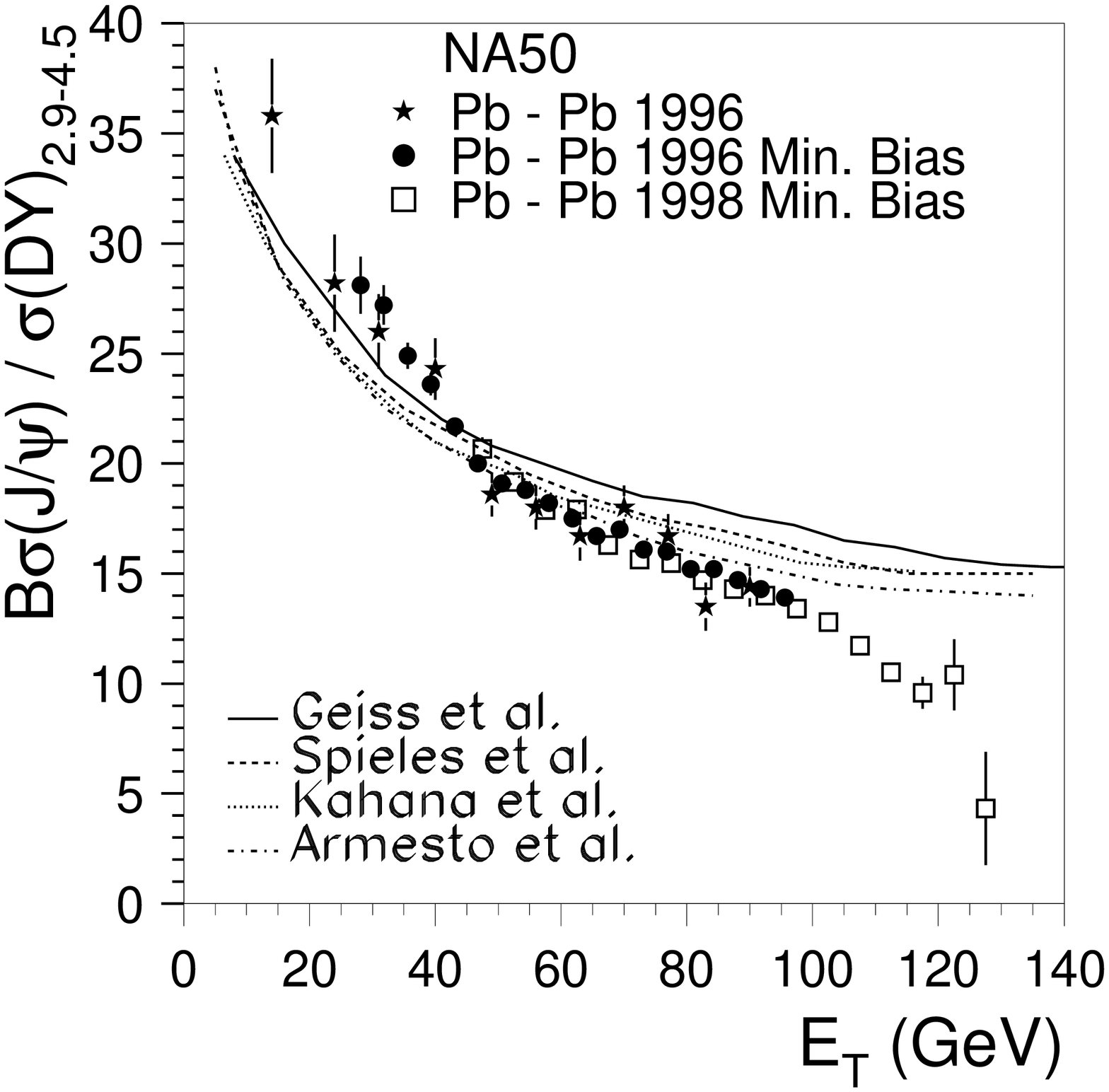}
%%%%%%% FIGURE 7b  CAPELLA CALCULATIONS
\includegraphics*[width=3.8cm]{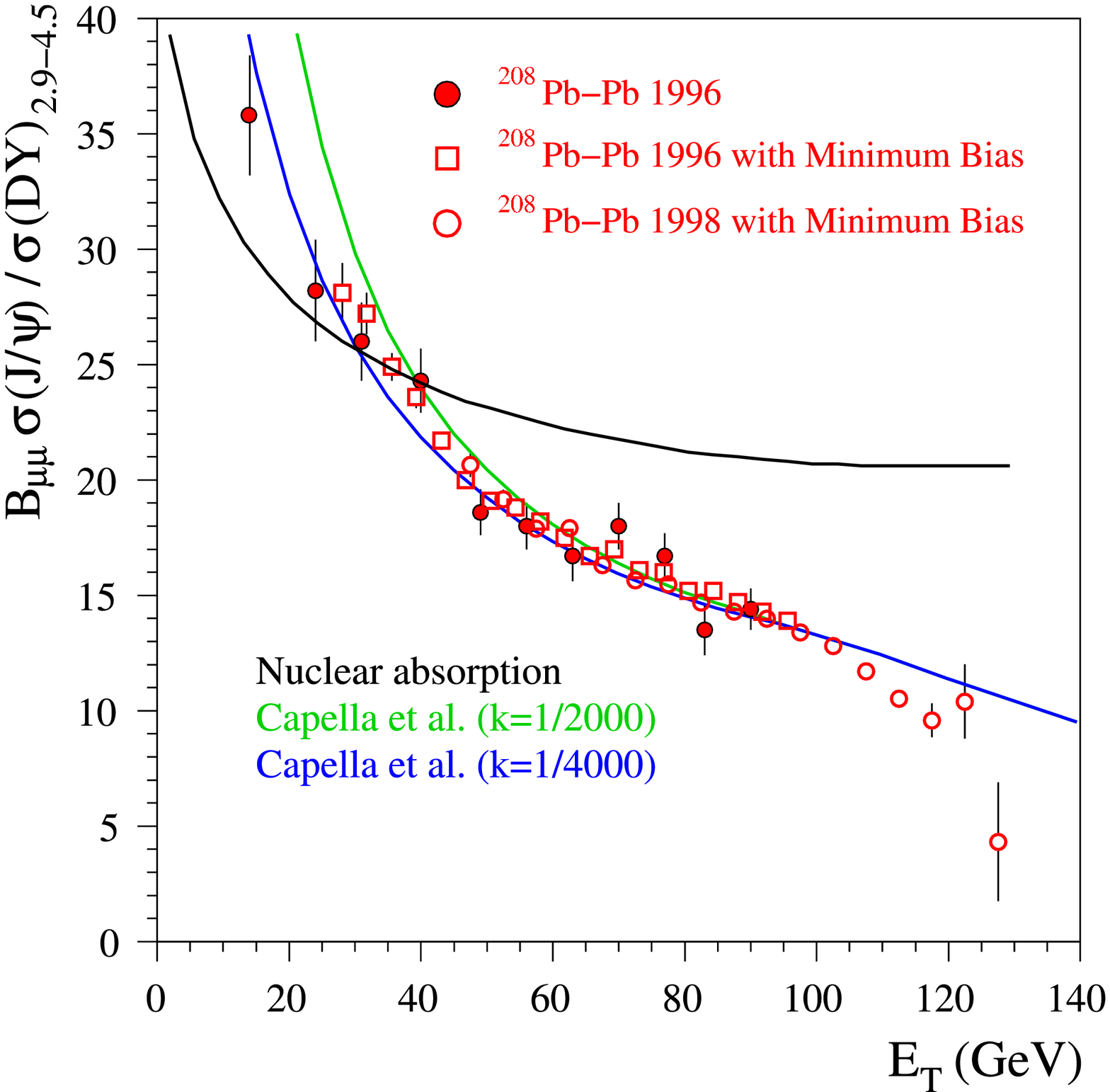}
%%%%%%% FIGURE 7c  QGP MODELS
\includegraphics*[width=3.8cm]{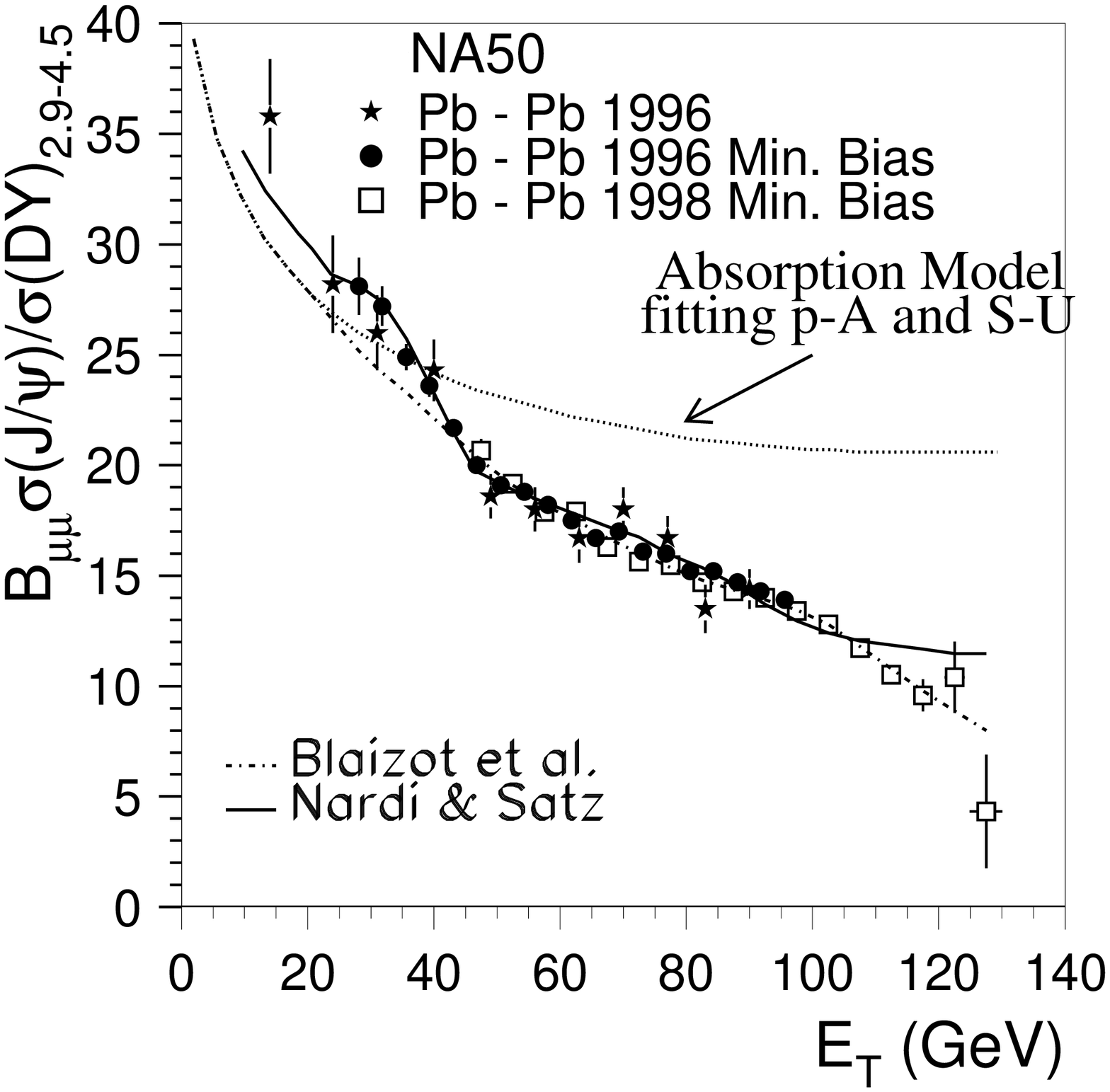}
\vspace{-.4cm}
\caption{J/$\psi$ over Drell-Yan cross section vs. E$_T$ measured by NA50 in Pb-Pb
         collisions at  $\sqrt{s_{NN}}$ = 17.2 GeV in comparison to various theoretical approaches 
         using conventional physics (left and middle panels) and quark matter
          formation (right panel) (taken from ~\protect\cite{bordalo-qm01}).}
\vspace{-0.9cm}
\end{figure}

\section{Suppression of large p$_T$ hadrons}
   
 With the energies available at RHIC, one order of magnitude higher than at the SPS,
new channels and probes become accessible for diagnostic of matter under extreme conditions.
In particular, the propagation in a dense medium of high p$_T$ partons resulting from hard
scattering and jet production in the initial phase of 
the collision, is an interesting question that has been extensively studied over the last 
decade \cite{wang-miklos-quenching}. Energy loss of the partons through gluon radiation
has been predicted as a possible signature of deconfined matter. The phenomenon, commonly referred
to as jet quenching, should manifest itself as a suppression of high  p$_T$ particles. 
Such a suppression has been observed already in the first low-luminosity RHIC run by the
two large experiments STAR \cite{harris-qm01} and PHENIX \cite{phenix-ppg003} and certainly constitutes 
the most interesting result from RHIC so far. The suppression is evidenced by plotting the so-called 
nuclear modification factor\cite{wang-wang} defined as the ratio of AA to pp 
p$_T$ spectra, scaled by the number of binary collisions:   

\[ R_{AA}(p_T) = \frac{d^2\sigma_{AA}/dp_T d\eta} { < N_{bin} > d^2\sigma_{pp}/ dp_T d\eta}  \]  

In the absence of any new physics this ratio should be equal to 1 at the high p$_T$
characteristic of hard processes. At low p$_T$,  dominated by soft processes which scale with the number of 
participants, the ratio is expected to be lower, e.g. for central collisions  
$R_{AA}(0) \approx 0.5 N_{part} / N_{bin} \approx 0.2$. The STAR and PHENIX results for central
Au-Au collisions at $\sqrt{s_{NN}}$ = 130~GeV are shown in
Figs. 8 and 9 respectively. In both cases the pp data at 130 GeV were derived from interpolation of pp data
\cite{ua1,isr,cdf} at lower and higher energies. 
%%%% FIGURES 8 AND 9 SIDE BY SIDE
%%%% FIGURE 8 STAR HIGH PT SUPPRESSION 
\begin{figure}[h]
\vspace{-0.2cm}
\begin{minipage}[t]{55mm}
\epsfig{file=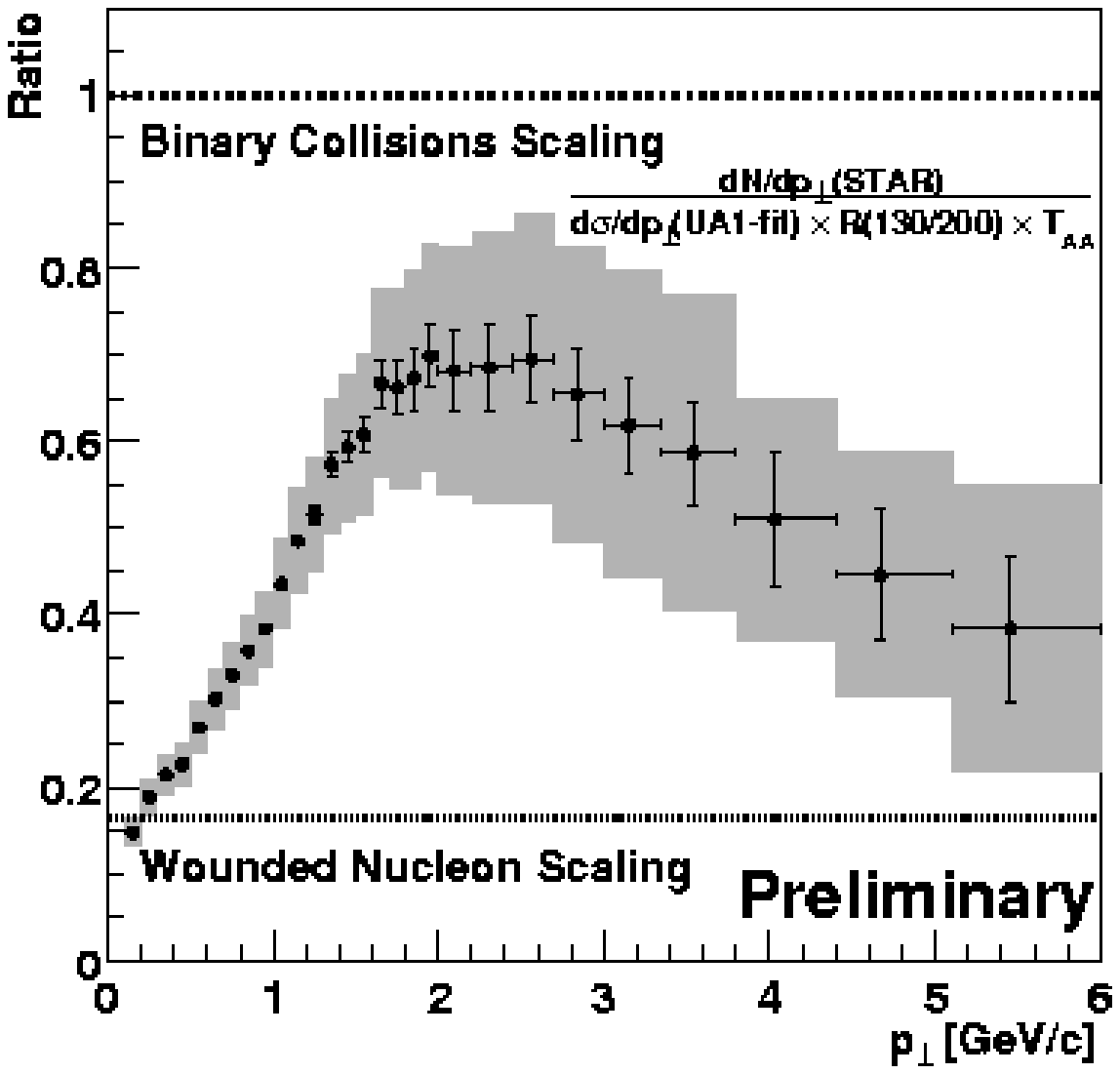,height=5.3cm}
\vspace{-0.6cm}
\caption{The ratio  R$_{AA}(p_T)$ for negative hadrons measured by STAR in
          Au-Au collisions at  $\sqrt{s_{NN}}$ = 130 GeV ~\protect\cite{harris-qm01}.}
\end{minipage}
\hspace{\fill}
%%%%%%% FIGURE 9.  PHENIX HIGH PT SUPPRESSION 
\begin{minipage}[ht]{55mm}
\vspace{-3.9cm}
\epsfig{file=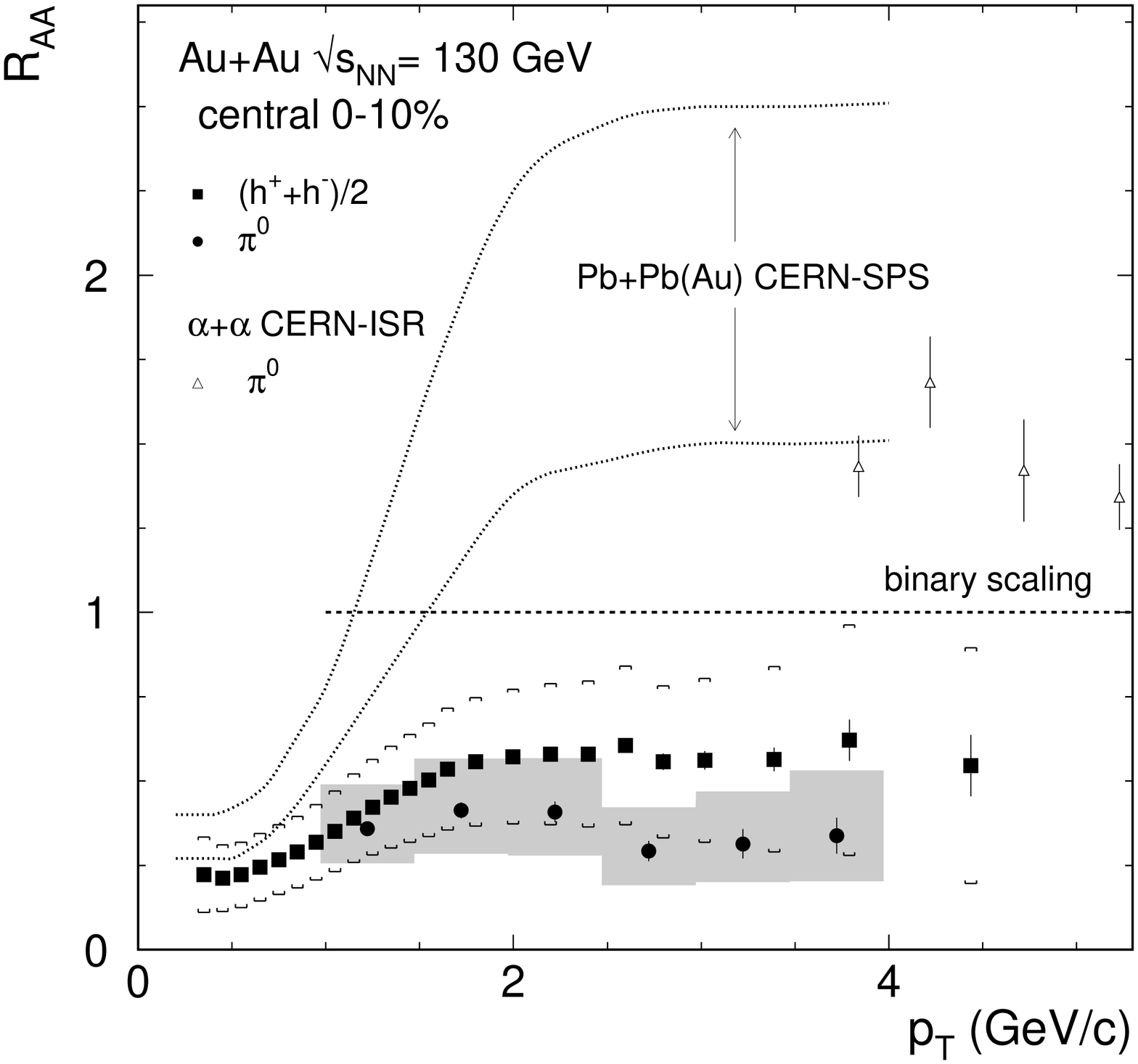,height=5.7cm}
\vspace{-1.1cm}
\caption {The ratio R$_{AA}(p_T)$ for negative hadrons and $\pi^o$ measured by PHENIX in
          Au-Au collisions at  $\sqrt{s_{NN}}$ = 130 GeV ~\protect\cite{phenix-ppg003}.}
\end{minipage}
\vspace{-0.5cm}
\end{figure}
 STAR shows the ratio for negative hadrons and PHENIX for negative hadrons as well as identified 
$\pi^o$. As expected, for low p$_T$ the ratio is small, $\approx 0.2$. It raises with  p$_T$, getting close to
1 and then falls off at higher p$_T$. The suppression appears to be stronger for
$\pi^o$ than for charged particles although both are consistent within their large systematic errors.  
A similar suppression pattern appears when the ratio of central to peripheral collisions, each divided 
by its corresponding value of $N_{bin}$, is plotted as function of p$_T$ \cite{phenix-ppg003}.
This behaviour is in marked contrast to the observations in Pb-Pb and Pb-Au collisions 
at the SPS (the solid lines in Fig. 10 represent the band of uncertainty of the CERN results)
where the ratio overshoots 1 and saturates at high  p$_T$, like in pA collisions \cite{wang-wang},  
an effect known as the Cronin effect \cite{cronin}. The high p$_T$ suppression observed at RHIC can 
be quantitatively reproduced in terms of 
jet quenching if an average energy loss of $dE/dx \approx$ 0.25 GeV/fm is assumed within a parton model
\cite{wang-qm01}. New RHIC data allowing to reach much higher  p$_T$ values together with 
reference data on pp and pA measured within the same apparatus will be very valuable to consolidate these
intriguing results.
  
\vspace{-0.3cm}
\section{Conclusion}
\vspace{-0.3cm}
   In a relatively short run, the RHIC experiments have produced an impressive amount of
results and much more is expected over the next years. The second RHIC run has just been 
completed with a recorded luminosity almost two orders of magnitude larger than in year-1.
This, together with the still ongoing yield of
interesting results from the SPS programme, places the field of relativistic heavy-ion collisions
at a very unique phase with exciting physics prospects.

\vspace{-0.3cm}
\section*{Acknowledgments}
\vspace{-0.3cm}
It is pleasure to acknowledge support from the MINERVA Foundation and the US-Israel Binational
Science Foundation.

\end{document}